 \documentclass[letter,useAMS,usenatbib]{mn2e}

\usepackage[dvips]{graphicx}
\usepackage{amssymb}
\usepackage{latexsym}
\usepackage{latexsym}
\usepackage{graphics}

\def\ha${H$\alpha$}
\def\hb${H$\beta$}
\def\hy{H$\gamma$}

\def\oii{[OII]$\lambda$3727}
\def\niib{[NII]$\lambda$6584}
\def\niia{[NII]$\lambda$6548}
\def\neiii{[NeIII]$\lambda$3869}
\def\oiiia{[OIII]$\lambda$4959}
\def\oiiib{[OIII]$\lambda$5007}

\def\siia{[SII]$\lambda$6716}
\def\siib{[SII]$\lambda$6730}

\def\apj{ApJ}
\def\apjs{ApJS}
\def\apjl{ApJL}
\def\aap{A\&A}	
\def\mnras{MNRAS}

\def\araa{ARAA}
\def\nat{Nature}

\title{The metallicity properties of zCOSMOS galaxies at 0.2$<$z$<$0.8}

\author[G. Cresci et al.]{
G. Cresci$^1$\thanks{E-mail:gcresci@arcetri.astro.it},
F. Mannucci$^{1}$,
V. Sommariva$^{1}$,
R. Maiolino$^{2}$,
A. Marconi$^{3}$,
M. Brusa$^{4}$\\
$^1$INAF - Osservatorio Astrofisico di Arcetri, 
   Largo E. Fermi 5, 50125 Firenze, Italy\\
$^2$INAF - Osservatorio Astronomico di Roma,
   via di Frascati 33, 00040 Monteporzio Catone, Italy \\
$^3$Universit\`a di Firenze - Dipartimento di Fisica e Astronomia,
   Largo E. Fermi 2, 50125 Firenze, Italy \\
$^4$Max Planck Institut f\"ur extraterrestrische Physik,
   Postfach 1312, 85741 Garching, Germany\\
}

\begin{document}

\date{}


\maketitle

\begin{abstract}
We study the metallicity properties of galaxies in the zCOSMOS sample between $0.2<z<0.8$. 
At $z<0.46$, where H$\alpha$ and [NII] are detected, we find the same dependence of metallicity on stellar mass 
and Star Formation Rate (SFR), the Fundamental Metallicity Relation, found by Mannucci et al. (\citealp{mannucci10}) 
in SDSS galaxies on a similar redshift range.
We extend this relation to higher redshift, $0.49<z<0.8$ where the $R_{23}$ metallicity index can be measured in our data, 
finding no evidence for evolution, and a metallicity scatter 
around the relation of about 0.16 dex. 
This result confirms, with a much higher level of significance with respect to previous works, the absence 
of evolution of the FMR during the last half of cosmic history.
\end{abstract}

%
\section{Introduction} \label{sec:intro}

The Fundamental Metallicity Relation (FMR) is a tight relation between
stellar mass, SFR, and metallicity, discovered by Mannucci et al. (\citealp{mannucci10}, hereafter M10) 
using local SDSS galaxies. The gas-phase metallicity 
of $\sim$140.000 galaxies with redshift between 0.07 and 0.3 
is found to depend not only on stellar mass, in the well-known 
mass-metallicity relation (Tremonti et al. \citealp{tremonti04}), but also on SFR, 
with more active galaxies showing lower metallicities. 
After considering this relation, the residual metallicity scatter 
is limited to $\sim$0.05 dex, which is compatible with the effect
of the uncertainties on mass, SFR, and metallicity. \\
The role of specific SFR (SSFR=SFR/Mass)) in relation with metallicity was first suggested by Ellison et al. (\citealp{ellison08}), who presented a mild ($\sim0.1$ dex) dependence of metallicity on SSFR when binning galaxies according to their stellar mass.
After M10, Lara-L\'opez et al. (\citealp{copioni}) discussed a plane in the same 3D space, and proposed to use it to derive the stellar mass of galaxies once their SFR and metallicity is known. 
The exact shape of the relation, especially at the high mass end, depends on the choice of the metallicity calibrations, on the assumptions about dust extinction and SFR, and on the extent of the aperture corrections allowed, as reported in Yates et al. (\citealp{yates11}).\\

Although the presence of the FMR is well established at low z, regardless of the metallicity calibration used, only a
limited amount of distant galaxies have been used for the study of the FMR at high redshifts.
The data available show that no evolution is present up to z=2.5, see Fig. 4 of M10),
i.e., galaxies of any mass and SFR follow the same FMR defined by local galaxies. The observed evolution of the mass-metallicity relation (e.g. Maiolino et al. \cite{maiolino08}, Mannucci et al. \cite{mannucci09}) is therefore due to a prospective effect, as high redshift galaxies usually have higher SFR and, as a consequence,  lower metallicities. 
Some evolution is observed at even higher redshifts, z$>$3, as discussed in 
M10 and Troncoso et al. (in preparation). \\
This results has been later confirmed by several groups. For example, Richard et al. (\citealp{richard11})
studied a sample of lensed galaxies at 1.4$<$z$<$5, which have SFR smaller than the 
other samples of un-lensed galaxies at these redshifts (Erb et al \citealp{erb06}, Maiolino et al. \citealp{maiolino08}, Mannucci et al. \citealp{mannucci09}). These galaxies show higher metallicity with respect to the mass-metallicity relation, exactly as expected by the FMR. Similarly, Nakajima et al. (\citealp{nakajima11}) presented the metallicities of stacked Ly$\alpha$ emitter at $z\sim2.2$, which have smaller SFR than other galaxies at similar redshift, but also higher metallicities according to the FMR. On the other hand, various highly star forming galaxies at high redshift are presented in the literature as very metal poor for their stellar mass. However, they result to be consistent with the FMR once their high SFR is considered (e.g. Erb et al. \citealp{erb10}, Kassin et al. \citealp{kassin11}). Finally, the use of the FMR allowed Mannucci et al. (\citealp{mannucci11})
to show that, when considering their high SFRs, the host galaxies of long-GRBs have the same 
metallicity distribution of field galaxies.\\

The high redshift galaxy samples used by M10 show a much larger 
spread in metallicity ($\sim$0.25 dex) around the FMR than local SDSS galaxies 
($\sim$0.05 dex). The spread is comparable with the much larger
uncertainties in SFR and metallicity, due to the generally lower signal-to-noise 
(S/N) ratio of the detected emission lines. This large spread could hide 
systematic trends, therefore it is important to test this results using larger samples of
galaxies. In this paper we use data from the zCOSMOS survey to study 
the evolution of the FMR up to $z\sim0.8$ with higher precision. This is fundamental to understand if the same long lasting equilibrium between star formation, infall of pristine gas and outflow of enriched material that seems to be responsible for the FMR in local galaxies (see e.g. Dav\'e et al. \citealp{dave11}, Torrey et al. \citealp{torrey11}) is also driving galaxy evolution at higher redshift, where infalls (Erb \citealp{erb08}, Mannucci et al. \citealp{mannucci09}, Cresci et al. \citealp{cresci10}) and outflows (e.g. Steidel et al. \citealp{steidel10}) are expected to be even more relevant. \\

The paper is organized as follows. In Sect. \ref{zCOSMOS} we briefly introduce the sample used, in Sect. \ref{SEDfitting} we present the method used to derive the stellar masses and in Sect. \ref{lines} we described how the line fluxes and the final sample selection were obtained. The measures of SFRs and extinctions are discussed in Sect. \ref{SFRAV}, while those of metallicities in Sect. \ref{compmet}. Our results are discussed in Sect. \ref{discussion}, while our conclusions follow in Sect. \ref{conclusions}.

\section{The zCOSMOS sample} \label{zCOSMOS}

zCOSMOS (Lilly et al. \citealp{lilly07}) is a redshift survey in the COSMOS field using the VIMOS multi-object spectrograph 
on ESO/VLT. It consists of two parts: the first, zCOSMOS-bright, is a redshift survey of $\sim20000$ I-band selected galaxies at redshifts $z < 1.2$, covering the approximately $1.7\ deg^2$ of the COSMOS field; the second part, zCOSMOS-deep, is observing $\sim10000$ galaxies selected through well-defined color selection criteria and at B<25 to be in the redshift range $1.5 < z < 3.0$. These observations are restricted to the central $1\ \textrm{deg}^2$ of the COSMOS field. 
Here we have used the spectra of the data release 2 of the zCOSMOS-bright survey, the so-called ``10k-bright'', available on the project web site since 1 Oct 2008\footnote{http://archive.eso.org/cms/eso-data/data-packages/zcosmos-data-release-dr2/}. These consist of $\sim10000$ spectra, obtained with $1''$ wide slits and the MR grism (R$\sim$600 with a wavelength range of approximately 5550 to 9650 \AA, sampled at roughly 2.5 \AA/pixel).  The details about the object selection, observations and data reduction can be found in Lilly et al. (\citealp{lilly09}).\\
In order to measure gas phase metallicity for our galaxies using strong optical emission lines, we need either [NII]$\lambda$6584 and H$\alpha$, or [OII]$\lambda$3727, [OIII]$\lambda$5007 and H$\beta$ visible in the wavelength range of the VIMOS spectrum. We therefore divided the obtained sample in two subsample at different redshifts: 2825 galaxies at $z<0.47$ have accessible H$\alpha$ emission and 3691 galaxies at $z>0.49$ are in the right redshift range to observe [OII] emission. \oiiib and H$\beta$ are observable for all these galaxies.\\

\section{Mass and continuum spectra from SED fitting} \label{SEDfitting}

To estimate the stellar masses we used 12 photometric bands: 
CFHT $u^{*}$ and $K_{s}$, Subaru $B_{j}$,  $V_{j}$, $g^{+}$,  $r^{+}$,
 $i^{+}$ and  $z^{+}$, UKIRT J (Capak et al. \citealp{cap07}, Ilbert et al. \citealp{ilb09}), and Spitzer IRAC at 3.6 $\mu$m,
4.5  $\mu$m and 5.8  $\mu$m (Frayer et al. \citealp{fray09}).\\
The photometry for all the optical and near-IR bands were measured over an aperture
of 3$''$ diameter, to minimize the effect of Point Spread Function variation from band to band. The derived fluxes are then corrected to total using the so called ``auto-offsets'', derived from the i-band image comparing the 3$''$ magnitude with the SExtractor total one (see Capak et al. \citealp{cap07}). 
The IRAC fluxes were measured in a circular aperture of radius 1.9$''$, converted to total fluxes using the correction factors derived by Surace et al. (\citealp{surace04}). We also used GALEX FUV and NUV photometry from Zamojski et al. (\citealp{zam07}), when available. A few galaxies show problematic 
values of the magnitude (Bolzonella et al. private communication), therefore GALEX photometry was 
only used when the reduced $\chi^{2}$ of SED fitting is below 2, for a total of 64 galaxies. \\

\begin{figure} 	\centering
	{\includegraphics[width=8cm]{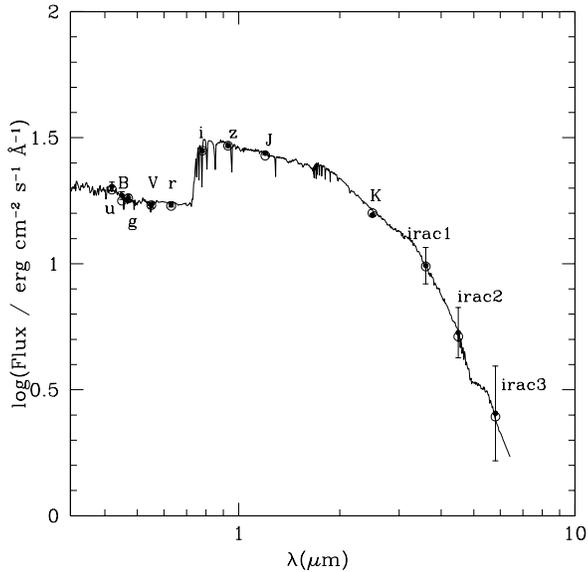}}
\caption{Example of SED fitting for one of the galaxies in the sample, ZC803435. The observed photometric data are shown as black solid points, while the expectations for each band based on the synthetic spectrum are shown as empty points.}
\label{sedfit}
\end{figure}

The stellar masses were obtained with a Spectral Energy Distribution (SED) fitting analysis, using the HyperZmass code (Pozzetti
et al. \citealp{poz07}, Pozzetti et al. \citealp{poz10}), a modified version of the HiperZ code (Bolzonella et al. \citealp{bol00}):  
the observed photometric SEDs are compared to those obtained 
from a set of reference synthetic spectra from stellar population models, 
and the best fitting SED is obtained using a $\chi^{2}$ minimization.
The best fitting spectrum and its normalization provide estimates of the star formation rate, extinction, stellar population age and 
galaxy stellar mass (as the integral of the SFR over the obtained age, correcting for mass losses during stellar evolution, 
see Pozzetti et al. \citealp{poz07}). In addition, the code makes available in output the best fitting synthetic stellar continuum 
for each galaxy, that has been used to correct for the Balmer absorptions (see Section \ref{lines}). \\
The reference spectral library was built with the Bruzual \& Charlot (\citealp{bc03}) code for spectral 
synthesis models, using the low resolution version with the Padova 1994 tracks.
The models were produced assuming a  Chabrier Initial Mass Function (IMF) 
(Chabrier et al. \citealp{cha03}) with an upper mass limit of 100$M_{\odot}$, the Calzetti (\citealp{cal00}) extinction law with A$_V$ varying between $0-4$ mag, 
and fixing the redshift to the spectroscopic value.
We used smooth exponentially decreasing Star Formation Histories, with time scale $\tau$ in the range $0.1-\infty$ and 
age $t$ between $0.1-20$ Gyr. An example of SED fitting for one of the galaxies of our sample is shown in Fig. \ref{sedfit}.\\
The stellar masses computed by Kauffmann et al. (\citealp{kau03a}) for the SDSS sample used by M10 to derive the FMR in the local universe
were instead computed using random secondary bursts in addition to the single one used here. 
Previous studies (e.g. Fontana et al. \citealp{fon04}, Pozzetti et al. \citealp{poz07}) 
have shown that random secondary
bursts may produce systematically larger masses up to about $0.2$ dex.
We therefore run several tests  adding random secondary burst to our synthetic spectra to check if this systematic difference in the stellar mass due to the use of a single exponential star formation history in our data is present and comparable to previous studies. In agreement to what is found in the literature, we found
that the stellar mass obtained with random secondary burst are $\sim 0.15-0.2$ dex larger than the ones obtained with a single burst. We therefore added 0.2 dex to the stellar masses obtained with a single burst for consistency with the SDSS stellar masses.\\
The stellar masses derived were also compared with the results obtained 
using the Maraston et al. (\citealp{M05}) spectral library, which include a more detailed treatment of the thermally pulsing asymptotic giant branch stars. These stars contribute significantly to the total luminosity in the $\sim0.2-2$ Gyr stellar population age. 
We found that the stellar masses obtained using the Maraston et al. (\citealp{M05}) library 
are lower on average by a factor of $\sim0.1$, as already mentioned in Bolzonella et al. (\citealp{bol10}).
To allow an easy comparison with the previous study in M10 for SDSS galaxies, we use in the following 
the results obtained with the Bruzual \& Charlot (\citealp{bc03}) library.\\

\section{Line fluxes} \label{lines}

\begin{figure}	\centering
	{\includegraphics[width=8.5cm]{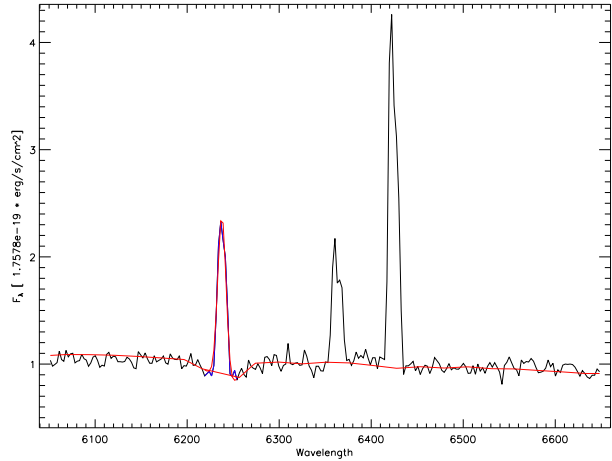}}
\caption{Example of the measurement of line fluxes by using the continuum obtained with SED fitting for one of the galaxies in the sample, ZC817676. The observed spectra is shown in black, while the continuum model and the best fitting Gaussian for the H$\beta$ line are shown in red. The [OIII] doublet is also visible in the spectral region shown.}
 \label{linefit}
\end{figure}

The extracted 1-D spectra provided by the zCOSMOS release were used to measure emission line fluxes. 
For each spectrum we fitted several emission lines, \oii, \neiii, \hy, H$\beta$, \oiiib, \oiiia, \niia, H$\alpha$, \niib, \siia, \siib,
when available in the accessible spectral range (see Section \ref{zCOSMOS}). The fit was performed using a custom IDL code, which automatically 
fits with a $\chi^2$ minimization a suitable stellar continuum and a Gaussian to each emission line.\\
To measure reliable fluxes, it is important to properly account for the stellar Balmer absorption, 
which in some galaxies can reach equivalent widths of $\sim5$\AA. The best fitting synthetic stellar continuum derived 
 from the SED fitting analysis of each galaxy (see Section \ref{SEDfitting}) is used to reproduce the absorption line profile and the 
 continuum shape. The synthetic stellar continuum is providing a very good representation of 
 the observed continuum in all the inspected spectra, providing a good estimate of the absorption underlying the Balmer emission lines: an example is shown in Fig. \ref{linefit}.\\
The continuum is fitted in a region of $\pm 300$ \AA~ around the expected position of each emission line. During the minimization, pixels in the data that consistently deviated more than 3$\sigma$ from the average were rejected, and were not used in the analysis. 
The uncertainties on the measured centroid position, velocity 
dispersion and flux are evaluated through Monte Carlo realizations, perturbing the input data assuming Gaussian uncertainties.\\

\section{Star Formation Rate and extinction} \label{SFRAV}

SFR can be derived from the H$\alpha$  and H$\beta$ lines, which provide a direct probe of the young massive stellar population. Similar to what has been done in M10 and many other authors, we have used the classical conversion factor by Kennicutt (\citealp{kennicutt98}), scaling down the results by a factor of 1.7 (Pozzetti et al. \citealp{poz07}) to convert them to the Chabrier (\citealp{cha03}) IMF. The conversion factor is based on the H$\alpha$ or H$\beta$ flux corrected for dust extinction, and therefore requires a good knowledge of the amount of extinction $A_{V}$ suffered by the emission lines. \\
For galaxies at $z<0.47$, where both H$\alpha$ and H$\beta$ are visible, the extinction can be derived directly from the Balmer decrement H$\alpha$/H$\beta$, assuming a Calzetti (\citealp{cal00}) extinction law, and the SFR from the H$\alpha$ extinction corrected flux. At $z>0.49$, where H$\alpha$ is not visible anymore in our data, we used the Balmer decrement of the fainter H$\gamma$/H$\beta$ for all the galaxies where S/N(H$\gamma$)$\geq5$, while the extinction $A_{V,SED}$ derived for the continuum by the SED fitting (see Section \ref{SEDfitting}) for the other galaxies. \\
Calzetti et al. (\citealp{cal00}) found evidence that the extinction derived from emission lines $A_{V,neb}$ in actively star forming galaxies is not equal but proportional to that measured from the continuum, with $A_{V,neb} = A_{V,SED}/0.44$ (see also e.g. F\"orster Schreiber et al. \citealp{natascha09}). We have compared $A_{V,neb}$ and $A_{V,SED}$ for all the galaxies where a Balmer decrement extinction measurement is available (either H$\alpha$/H$\beta$ or H$\gamma$/H$\beta$), finding that the extra attenuation of the emission lines is confirmed by our data: our best fitting relation is $A_{V,SED}=(0.42\pm0.01)\cdot A_{V,neb}$. Given the small differences, we adopt  the relation $A_{V,neb} = A_{V,SED}/0.44$ of Calzetti et al. (\citealp{cal00}) to all the objects in the higher redshift bin with no H$\gamma$ flux, in order to ensure a direct comparison of our results with previous works.\\


\section{Metallicity} \label{compmet}

Given the large variety of different strong line metallicity calibrations used in the literature 
(see e.g. Pagel et al. \citealp{pagel79}, Pettini \& Pagel \citealp{pp04};
Kewley \& Dopita \citealp{kewley02}; Dopita et al. \citealp{dopita06}; Tremonti et al. \citealp{tremonti04}; Nagao et al. \citealp{nagao06}), which give very different results for a given line ratio, it is mandatory to study the evolution of the FMR using the same calibration used in M10. Consistently with that work, gas-phase metallicities were computed using the calibrations obtained by Nagao et al. (\citealp{nagao06}) and Maiolino et al. (\citealp{maiolino08}), to allow a proper comparison with local SDSS galaxies. 
These ratios have been empirically calibrated at low metallicities ($12+log(O/H)<8.4$) using direct $T_e$ measurements in local metal poor galaxies and HII regions, and are based on the photoionization models by Kewley \& Dopita \citealp{kewley02} (hereafter KD02) at high metallicity. The metallicity is therefore provided on the KD02 scale, while a conversion to other metallicity systems can be found in Kewley \& Ellison (\citealp{kewley08}). The calibration proposed in KD02 is dependent on the [NII]/[OII] ratio, but the use of a secondary element as Nitrogen is of some concern. This issue will be addressed in detail by Maiolino et al. (in preparation, see also P\'erez-Montero \& Contini \citealp{perez09}, Torres-Papaqui et al. \citealp{tp11}).\\

Two different measurements of metallicity are available for the two redshift ranges. At $z<0.47$, where H$\alpha$ is still visible, the [NII]/H$\alpha$ ratio is used. At $z>0.49$, where [OII] emission is in the accessible wavelength range, the measurement is instead based on the $R_{23}$ quantity, defined as $R_{23}$=(\oii+\oiiia+\oiiib)/H$\beta$. In this latter case, as the lines involved are significantly separated in wavelength, the extinction correction discussed is Section \ref{SFRAV} has to be applied to the observed line fluxes. Moreover, as $R_{23}$ has two possible metallicity solutions for a given line ratio, other diagnostics have to be used to remove the degeneracy and distinguish between the two branches at low and high metallicity. \\

We use for this task both the [OIII]/[OII]  and [NeIII]/[OII] line ratios, which have a monotonic dependence on metallicity, but are also affected by a larger dispersion due to their additional dependence on the ionization parameter (see Nagao et al. \citealp{nagao06}). In spite of their larger scatter, low and high metallicity galaxies are reasonably well separated in both diagrams (see e.g. Fig. 5 in Maiolino et al. \citealp{maiolino08}). \\
We therefore assigned all galaxies with log([OIII]/[OII])$<0.45$ and log([NeIII]/[OII])$<-0.6$ to the high metallicity $R_{23}$ branch, while all the others to the low metallicity one. The two independent branch indicators used agree for all the galaxies in our sample, supporting our method to discriminate between branches. \\
To further check the reliability of the branch selection done, we add to the galaxies used in Nagao et al. (\citealp{nagao06}) and Maiolino et al. (\citealp{maiolino08}) the sample of $\sim 600$ SDSS with direct $T_e$ measurement of Kniazev et al. (\citealp{kniazev04}). We find that their galaxies with $S/N>5$ on the [OIII]$\lambda$4363 line (used for $T_e$ measurement, see e.g. Izotov et al. \citealp{izotov06}) follow our criteria on [OIII]/[OII] to discriminate between the two branches. This is also confirmed by a larger sample of galaxies with $T_e$ measurement that will be presented in a forthcoming paper (Maiolino et al. in preparation).\\

The uncertainties on the gas phase metallicities are of the order of $0.1-0.2$ dex, and are dominated by the spread in the calibrations themselves. The uncertainties in the higher redshift bin tend to be larger when the metallicity has values 12 + log(O/H) $\sim 8$, where $R_{23}$ has a maximum and a small uncertainty in the line ratio produces a large uncertainty in metallicity. In any case, we note that just very few galaxies in our sample are in the low metallicity region where $R_{23}$ is flat (between 12+log(O/H)$\sim7.7-8.4$, see Fig. \ref{massmet}) \\


\section{Sample selection}

The results of the SED and emission line fitting described in Sect. \ref{SEDfitting} and \ref{lines} above were combined and used to define the two final subsamples (galaxies at $z<0.47$ with accessible H$\alpha$ emission and galaxies at $z>0.49$ with [OII] emission). We applied to the initial sample discussed in Section \ref{zCOSMOS} the following selections:

\begin{enumerate}

\item each zCOSMOS spectrum is assigned to a confidence class according on the reliability of the final redshift. We selected all the objects with confidence classes 3 and 4 (respectively, ``secure'' and ``very secure'' redshifts), and class 2.5 (redshift with reliability larger than 85\% and photometric redshift 
in good agreement with the spectroscopic redshift). We also added all the objects in classes 23 and 24, i.e. secondary targets detected in a slit positioned on another target, but with secure and very secure redshifts. We further excluded from our sample duplicate objects. Galaxies classified as broad-line AGN in the zCOSMOS catalog (class $>10$) or stars (class $*.4$) are automatically excluded from our selection. 

\item as our two subsamples may be still contaminated by type-2 narrow line AGNs, or with by AGN with nuclear emission diluted in the host galaxy light, we applied two additional filterings. First, we excluded objects associated to X-rays selected AGN from the XMM-COSMOS catalog with rest-frame hard X-rays luminosity in excess than $2\cdot 10^{42}\ erg\ s^{-1}$, typical of AGNs (see Brusa et al. \citealp{marcella10} for more details). Second, to further exclude Compton Thick and Low Luminosity AGN below the XMM-COSMOS sensitivity, we used standard BPT selections based on emission line ratios. Following Bongiorno et al. (\citealp{angela10}), we used the [OIII]/H$\beta$ versus [NII]/H$\alpha$ diagnostic diagram  (Kauffmann et al. \citealp{kau03b}) for galaxies at $z<0.46$, and the [OIII]/H$\beta$ versus [OII]/H$\beta$ diagram (Lamareille et al. \citealp{lam04}) for galaxies at $z>0.49$;

\item the velocity shift between each fitted emission line centroid and the zCOSMOS redshift 
must be below 280 km/sec, and the line width must be between 110 km/sec and 290 km/sec,
otherwise the fit is considered unreliable and the galaxy is excluded. 
For this selection we considered all the main emission lines: H$\beta$\ and \oiiib\ for 
all the galaxies, H$\alpha$\ and \niia\ for galaxies at low redshift (z$<$0.46), \oii~ 
for galaxies at high redshift (z$>$0.49);

\item the same lines must not correspond to bright sky lines, or to spectral regions badly affected by fringing ($\lambda>8000$ \AA);

\item the reduced $\chi^2$ of SED fitting must be below 5, and the final 68\% uncertainty on mass must be below 0.30 dex;

\item a threshold on the S/N of H$\alpha$ and H$\beta$ is also applied to have reliable and unbiased metallicity measurements.
As line ratios depend strongly on metallicity (see, for example, Maiolino et al. \citealp{maiolino08}),
applying the same thresholds to all  relevant lines or to their ratios would introduce severe biases in metallicity (see M10 for a detailed discussion). 
To avoid these biases, and following what has been done for M10, we only apply high S/N 
thresholds to the lines which are diagnostics of SFR  like H$\alpha$ and H$\beta$. The actual threshold values depend on the redshift and stellar mass:
at z$<$0.46 we used S/N(H$\alpha$)=15 for $log(M_*/M_{\odot})\geq9.4$, and S/N(H$\alpha$)=30 for $log(M_*/M_{\odot})<9.4$. At $z>0.49$ we used S/N(H$\beta$)=8 for $log(M_*/M_{\odot})\leq9.8$, and S/N(H$\beta$)=20 for $log(M_*/M_{\odot})>9.8$.
With these choices all the other relevant lines are detected with adequate S/N without additional, metal dependent selections. 
Indeed, in our lower redshift bin biases on metallicity are important mainly at low stellar masses: for example, at $log(M/M_{\odot})\sim9$, where metallicities are 12+log(O/H)$\sim8.5$, [NII] is only $\sim0.1$ H$\alpha$, while at $log(M/M_{\odot})\sim10.2$, where the metallicity is as high as 12+log(O/H)$\sim9$,  [NII] is $\sim0.4$ H$\alpha$, and a lower S/N threshold on H$\alpha$ may be applied to maximize the number of galaxies in our sample. In the higher redshift bin, on the other hand, the situation is reversed and [OIII], which is only $\sim0.4$ H$\beta$ in the high mass bin, increases to $\sim3$ H$\beta$ at the lower masses and  metallicities. \\
These considerations indicate that our choice of  S/N thresholds on H$\alpha$ and H$\beta$ lines that depends on mass in both redshift bins maximize the number of galaxies selected without introducing metallicity biases;

\item following M10, we checked the metallicity derived from \niia/H$\alpha$ for galaxies at $z<0.46$ using an independent metallicity indicator. As [OII] is not available in this redshift range, we used [OIII]/H$\beta$, which has properties similar to $R_{23}$ with just a slightly larger scatter: we selected only galaxies where the values of metallicity obtained with the two different indicators differ less that 0.25 dex. At $z>0.49$, instead, no additional independent metallicity tracer with low enough scatter is available, and no additional selection is applied;

\end{enumerate}

\begin{figure} 
	\centering
	{\includegraphics[width=8cm]{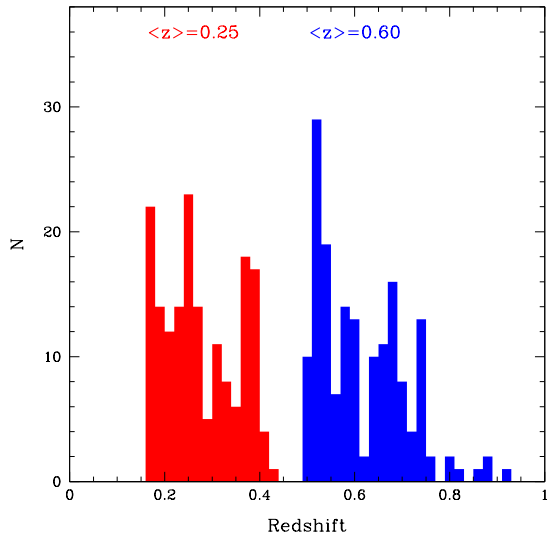}}
\caption{Redshift distribution of the two final galaxy subsamples (see text). The galaxies at $z<0.47$ with visible H$\alpha$ emission are shown in red, and galaxies at $z>0.49$ with [OII] emission are shown in blue.}
\label{zdistr}
\end{figure}

\begin{figure*}
	\centering
	{\includegraphics[angle=-90,width=15cm]{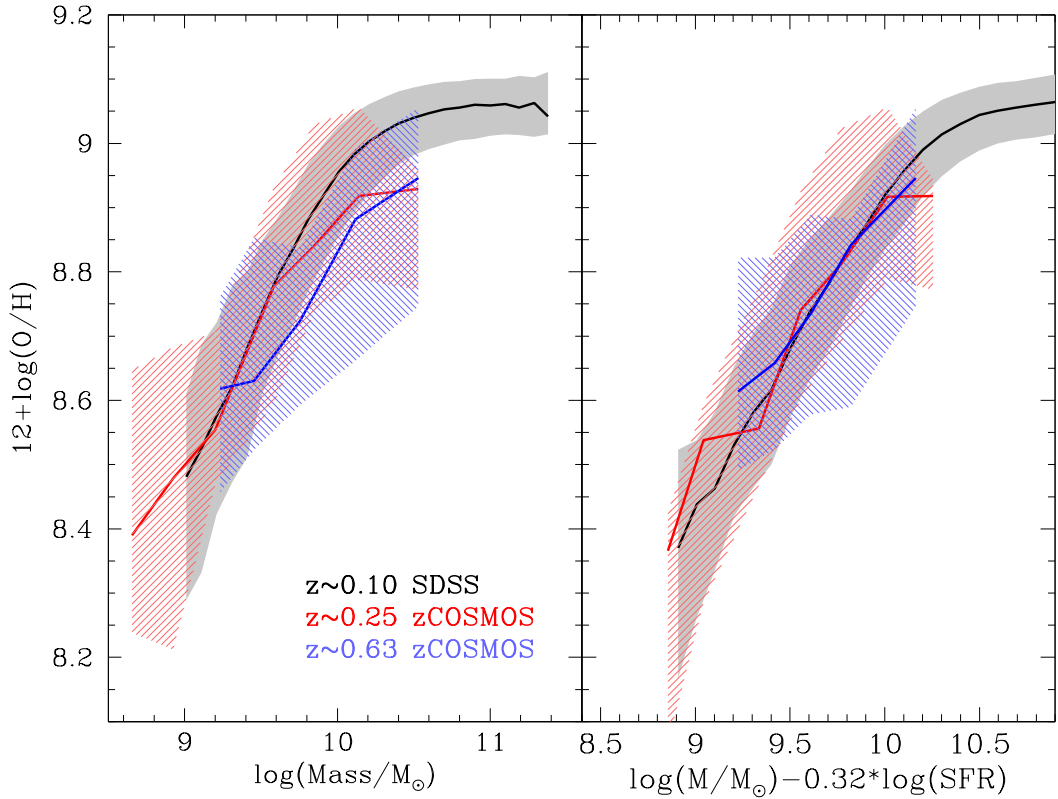}}
\caption{The mass metallicity relation (left panel) and the $\mu_{0.32}$ projection of the Fundamental Metallicity Relation (right panel) for the $z<0.46$ (red solid lines) and the $z>0.49$ (blue solid lines) zCOSMOS galaxies. The shaded areas contain galaxies between the 16th and the 84th percentile rank. The two high redshift relations are compared with the local one, as defined by the SDSS galaxies from M10 (black solid lines). It is evident an evolution of the mass metallicity relation at $z>0.49$, while the corresponding projections on the FMR plane shows no evidence of evolution even at the higher redshift.}
\label{massmet}
\end{figure*}

After applying all these selections, we obtain 169 galaxies at $z<0.47$ ($<z>=0.25$) and 165 galaxies at $z>0.49$ ($<z>=0.60)$. 
The redshift distribution of the two final subsample is shown in Fig. \ref{zdistr}.

\section{Results} \label{discussion}

\begin{figure} 
	\centering
	{\includegraphics[width=8cm]{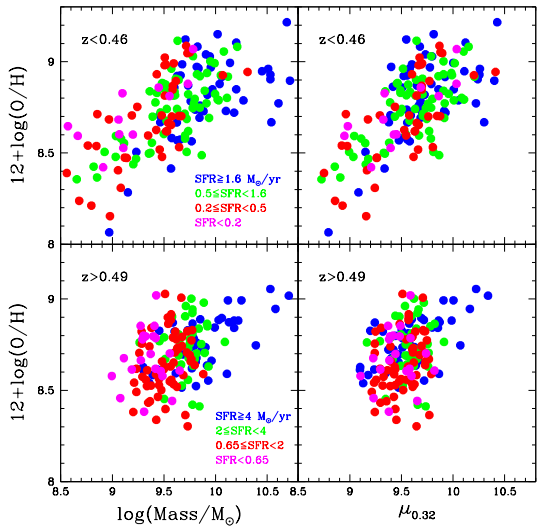}}
\caption{The mass metallicity relation (left panels) and the $\mu{0.32}$ projection of the FMR (right panels) for the individual galaxies of the sample. The lower redshift bin ($z<0.46$) is shown in the upper panels, while the higher redshift bin ($z>0.49$) in the lower panels. The galaxies are color coded as a function of their SFR, with decreasing SFR from blue to magenta. While a trend for lower metallicity at higher SFR and fixed mass is present in the mass metallicity relation at both redshifts, the projection removes such effects, showing that the galaxies up to $z\sim0.8$ populate a 3D plane defined by mass, metallicity and SFR as local SDSS galaxies. }
\label{scatter}
\end{figure}

The Mass-Metallicity relations for the two zCOSMOS samples at $z\sim0.25$ and $z\sim0.60$ are shown in Fig. \ref{massmet} (left panels). As expected, the Mass-Metallicity relation for the $z\sim0.25$ sample is comparable to the relation derived by M10 for SDSS galaxies, as the SDSS sample used by M10 is including galaxies up to $z\lesssim0.3$.\\
In the higher redshift bin, consistently with what was observed in other samples at comparable redshift (e.g. Savaglio et al. \citealp{savaglio05}, Cowie \& Barger \citealp{cb08}, Rodrigues et al. \citealp{rodrigues08}, Lamareille et al. \citealp{lam09}, P\'erez-Motero et al. \citealp{pm09}, Zahid et al. \citealp{zahid11}), we find significant evolution of the mass metallicity relation, of $\sim0.1$ dex.
Moreover, if we plot the single galaxies on the mass metallicity relation  color coded as a function of their SFR, as shown in Fig. \ref{scatter} (left panels) for both redshift bins, there is a trend for more star forming galaxies to have lower metallicity at a given stellar mass, as found by M10 for the SDSS sample. \\

M10, in fact, defined the FMR adding the SFR of the galaxies as a third parameter in the mass metallicity relation, showing that SDSS galaxies with lower metallicity also have higher SFR at a given mass, and that roughly half of the scatter in the relation was due to the different individual SFR of the galaxies. They also showed that the  metallicity scatter of local galaxies could be minimized plotting them on a particular projection of the FMR, introducing a new quantity $\mu_{0.32}$ defined as:
\begin{equation}
	\mu_{0.32}=log(M/M_{\odot})-0.32\cdot log(SFR)
\end{equation}
As shown in the right panel of Fig. \ref{massmet}, no evolution is observed even in the higher redshift bin in the $\mu_{0.32}$ plane, once the increased SFR of the zCOSMOS galaxies at high z with respect to the local SDSS galaxies is accounted for. Moreover, the projection clearly removes any dependence of the mass-metallicity scatter from the SFR of the galaxies (see Fig. \ref{scatter}, right panels). This shows that the star forming galaxies up to $z\sim0.8$ populate the same tight surface in the 3D space defined by their mass, metallicity and SFR as demonstrated by M10 for local SDSS galaxies.\\

\begin{figure}
	\centering
	{\includegraphics[width=8cm,keepaspectratio]{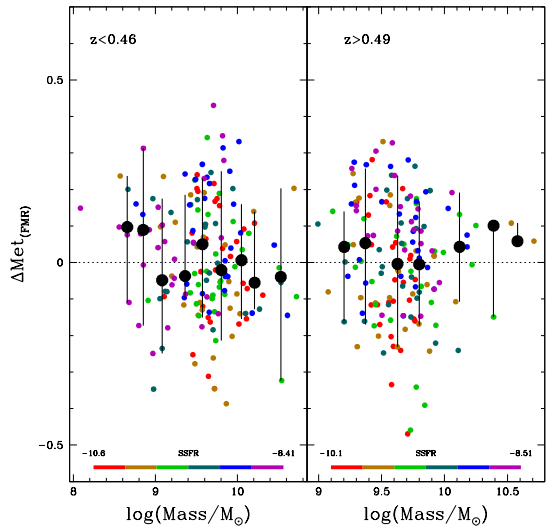}}
\caption{Difference from the FMR surface for the zCOSMOS galaxies at $z<0.46$ (left panel) and $z>0.49$ (right panel) as a function of stellar mass. The single points are shown color coded as a function of their specific SFR, while their average in each bin along with the 1$\sigma$ scatter is shown in black. The galaxies in both redshift bins are consistent with no differences from the FMR, with an average offset of $9\cdot10^{-3}\pm0.17$ at $z<0.46$ and $9\cdot10^{-3}\pm0.16$ at $z>0.49$. No residual dependencies from the SSFR is found.}
 \label{plotfmr}
\end{figure}

To better check the presence of any evolution of the FMR with redshift, we compute the expected metallicity of each galaxy, given its stellar mass and SFR, according to the 3D plane defined by local SDSS in  M10. We then plot in Fig. \ref{plotfmr} the difference of the observed value with the expected metallicity. The galaxies in both redshift bins are consistent with no differences from the FMR, with an average offset of $9\cdot10^{-3}\pm0.17$ at $z<0.46$ and $9\cdot10^{-3}\pm0.16$ at $z>0.49$, confirming with a much larger sample the no evolution of the FMR up to $z\sim0.8$ suggested by M10.\\

The larger observed scatter in the zCOSMOS data with respect to the SDSS galaxies of M10 is partly due to instrumental effects. In fact, the VIMOS spectra used here are affected by significant fringing beyond $\sim8000$ \AA~ (see e.g. Le F{\`e}vre et al. \citealp{lefevre05}). Although we excluded galaxies with emission lines in particularly bad part of the spectrum, our line fitting quality is anyway degraded redward of $\sim8000$ \AA. In fact, if we exclude the galaxies with emission lines contaminated by fringing (i.e. galaxies at $z>0.22$ for our lower redshift bin and objects at $z>0.6$ for the higher redshift bin), the scatter is significantly reduced to 0.11 dex and 0.14 dex respectively for the two subsamples. 
Moreover, the scatter in the local FMR based on SDSS galaxies depends significantly on mass and SFR, increasing at higher SFR and lower stellar mass (see Table 1 in M10). Therefore, the scatter derived for the zCOSMOS objects has to be compared with that obtained for SDSS galaxies with comparable stellar mass and SFR only: the $z=0$ scatter corresponding to the lower redshift zCOSMOS subsample galaxy properties is 0.10 dex, while the one relative to higher redshift one is 0.11 dex, closer to the fringing free scatter observed in our data. The remaining difference in the higher redshift bin is probably due to the single metallicity indicator used for those galaxies, while in M10 the abundances were derived using both $R_{23}$ and H$\alpha$ for all the galaxies.\\ 

\begin{figure}
	\centering
	{\includegraphics[width=8cm,keepaspectratio]{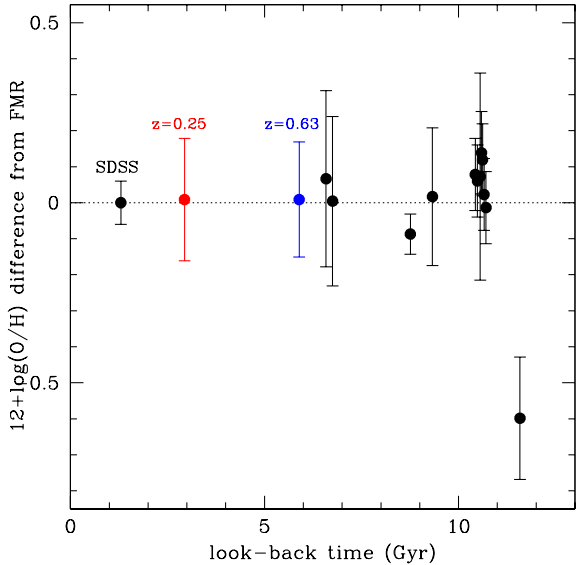}}
	\caption{Redshift evolution from the FMR plane defined by local SDSS galaxies for zCOSMOS galaxies (this work, red and blue) and other high-z sample in the literature ($0.5<z<0.9$: Savaglio et al. 2005; $1.0<z<1.6$: Shapley et al. 2005, Liu et al. 2008, Wright et al. 2009, Epinat et al. 2009; $2.0<z<2.5$: Law et al. 2009, F\"orster Schreiber et al. 2009; $z\sim3.3$: Maiolino et al. 2008, Mannucci et al. 2009). All the galaxy samples up to z=2.5 are consistent with no evolution of the FMR defined locally, while metallicities lower by $\sim0.6$ dex are observed at $z=3.3$. While each black point at $z>0.7$ is the average of $\sim 20$ galaxies or less, the points at $z\sim0.25$ and $z\sim0.60$ comprise $\sim160$ galaxies each.}
\label{zevol}
\end{figure}

The redshift evolution of the FMR based on the zCOSMOS data presented in this work and on higher redshift data from the literature compiled by M10 is shown in Fig. \ref{zevol}.  All the galaxy samples up to z=2.5 are consistent with no evolution of the FMR defined locally, while metallicities lower by $\sim0.6$ dex are observed at $z=3.3$ (see M10). As the data points at $z>1$ are based on few massive galaxies, the larger samples at high redshift presented here at $z\sim0.25$ and $z\sim0.63$ provide a stronger evidence for the no-evolution of the FMR with redshift.

\section{Conclusions} \label{conclusions}

In this paper we have studied the gas phase metallicity of a sample of more than 300 star forming galaxies between $0.2 \lesssim z \lesssim 0.8$ from the zCOSMOS survey. We have measured the gas phase metallicity using strong optical line ratios, \niia/H$\alpha$ for galaxies at $z<0.46$ and $R_{23}$ for galaxies at $z>0.49$. \\
We find that the metallicity of the galaxies is tightly related to their stellar mass, as for local galaxies. However, a clear evolution of the mass metallicity relation of $\sim0.1$ dex with respect to the local one is observed for the galaxies with $z>0.49$. 
We also found a dependence of the scatter observed around the mass metallicity of our sample on the SFR, showing that star forming galaxies up to $z\sim0.8$ are defining a 3D plane between stellar mass, metallicity and SFR, as local SDSS galaxies. Moreover, we have shown that projecting the galaxies on the FMR plane using the $\mu_{0.32}$ quantity successfully removes the dependence of the scatter by the SFR of the galaxies, and that on this projection the higher redshift zCOSMOS galaxies follow the relation defined by local SDSS galaxies with no evidences of evolution.\\
The lack of evolution of the FMR at $0.2<z<0.8$ is fully confirmed for the zCOSMOS sample, as the expected metallicities computed according to the local FMR using the mass and SFR of the galaxies are consistent with the observed ones: the average offset is $9\cdot10^{-3}\pm0.17$ at $z<0.46$ and $9\cdot10^{-3}\pm0.16$ at $z>0.49$. The observed scatter is somehow larger than the one measured for SDSS galaxies, although the difference is probably due to VIMOS instrumental effects and to the lower number of independent metallicity indicators available.\\
The results presented confirm with much higher level of significance that the same physical processes are probably responsible for the SFR, metal enrichment and gas recycling through infall and outflows in the local and higher redshift Universe, as already suggested by M10. Moreover, they support the idea that the observed evolution of the mass-metallicity relation, at least up to $z\sim0.8$, is simply due to more severe selection biases at high redshift as well as to the increase of the average SFR, which results in sampling different parts of the same FMR at different cosmic epochs.


\vspace{1cm}

\noindent
{\bf Acknowledgments}\\

We wish to thank the zCOSMOS project for making the data available to the community, and L. Pozzetti e M. Bolzonella for assistance in computing stellar masses. We  also wish to thank  M. Mignoli for support in understanding zCOSMOS spectra, and C. Maier for the stimulating discussion about metallicity measurements in this sample. GC and FM acknowledge financial support from the Italian Space Agency through ASI-INAF grant I/009/10/0. FM acknowledges additional support from ASI-INAF grant I/016/07/0, and from PRIN-INAF 2008. This work is based on observations made with ESO Telescopes at Paranal Observatories under programme ID 175.A-0839839, and in part on observations obtained with XMM-Newton, an ESA Science Mission with instruments and contributions directly founded by ESA Member States and the USA (NASA).


\end{document}